\documentclass[preprint]{iucr}              % DO NOT DELETE THIS LINE
\usepackage{graphicx}% Include figure files

     \journalcode{M} % Indicate the journal to which submitted
\begin{document}     % DO NOT DELETE THIS LINE

     %-------------------------------------------------------------------------
     % The introductory (header) part of the paper
     %-------------------------------------------------------------------------

     % The title of the paper. Use \shorttitle to indicate an abbreviated title
     % for use in running heads (you will need to uncomment it).

\title{Orthorhombic charge density wave on the
tetragonal lattice of EuAl$_4$}

\shorttitle{Orthorhombic CDW in EuAl$_4$}

     % Authors' names and addresses. Use \cauthor for the main (contact) author.
     % Use \author for all other authors. Use \aff for authors' affiliations.
     % Use lower-case letters in square brackets to link authors to their
     % affiliations; if there is only one affiliation address, remove the [a].

\cauthor[a,b]{Sitaram}{Ramakrishnan}{niranj002@gmail.com}{}
\author[a]{Surya Rohith}{Kotla}
\author[a]{Toms}{Rekis}
\author[a,c]{Jin-Ke}{Bao}
\author[a]{Claudio}{Eisele}
\author[d]{Leila}{Noohinejad}
\author[d]{Martin}{Tolkiehn}
\author[d,e]{Carsten}{Paulmann}
\author[f]{Birender}{Singh}
\author[f]{Rahul}{Verma}
\author[f]{Biplab}{Bag}
\author[f]{Ruta}{Kulkarni}
\author[f]{Arumugam}{Thamizhavel}
\cauthor[f]{Bahadur}{Singh}{bahadur.singh@tifr.res.in}{}
\cauthor[f]{Srinivasan}{Ramakrishnan}{ramky07@gmail.com}{}
\cauthor[a]{Sander}{van Smaalen}{smash@uni-bayreuth.de}{}

\aff[a]{Laboratory of Crystallography, University
of Bayreuth, 95447 Bayreuth \country{Germany}}

\aff[b]{Department of Quantum Matter, AdSM,
Hiroshima University,
739-8530, Higashi-Hiroshima  \country{Japan}}

\aff[c]{Department of Physics,
Materials Genome Institute and International Center
for Quantum and Molecular Structures,
Shanghai University, Shanghai 200444, \country{P. R. China}}

\aff[d]{P24, PETRA III,
Deutsches Elektronen-Synchrotron DESY,
Notkestr. 85, 22607 Hamburg \country{Germany}}

\aff[e]{Mineralogisch-Petrographisches Institut,%
Universit\"{a}t Hamburg, 20146 Hamburg \country{Germany}}

\aff[f]{Department of Condensed Matter Physics
and Materials Science,Tata Institute of Fundamental Research,
Mumbai 400005 \country{India}}

     % Use \shortauthor to indicate an abbreviated author list for use in
     % running heads (you will need to uncomment it).

\shortauthor{Ramakrishnan, Kotla, Rekis \textit{et al.}}

     % Use \vita if required to give biographical details (for authors of
     % invited review papers only). Uncomment it.


\maketitle       % DO NOT DELETE THIS LINE

\begin{synopsis}
The incommensurate charge-density-wave of EuAl$_4$
below $T_{CDW} = 145$\ K is found to possess orthorhombic
symmetry, despite an average crystal structure
that remains tetragonal in very good approximation.
This finding has ramifications for the interpretation
of all physical properties of EuAl$_4$,
in particular its multiple magnetic transitions.
\end{synopsis}

\begin{abstract}
EuAl$_4$ possesses the BaAl$_4$ crystal structure type
with tetragonal symmetry $I4/mmm$.
It undergoes a charge-density-wave (CDW)
transition at $T_{CDW} = 145$\ K
and it features four consecutive antiferromagnetic
phase transitions below 16\ K.
Here, we use single-crystal x-ray diffraction
to determine incommensurately
modulated crystal structure of EuAl$_4$ in its
CDW state.
The CDW is shown to be incommensurate with
modulation wave vector
$\mathbf{q} = (0, 0, 0.1781(3))$ at 70\ K.
The symmetry of the incommensurately modulated
crystal structure is orthorhombic with
superspace group $Fmmm(00\sigma)s00$,
where $Fmmm$ is a subgroup of $I4/mmm$ of index 2.
Both the lattice and the atomic coordinates
of the basic structure remain tetragonal.
Symmetry breaking is entirely due to the
modulation wave, where atoms Eu and Al1
have displacements exclusively along
$\mathbf{a}$, while the fourfold rotation
would require equal displacement amplitudes along
$\mathbf{a}$ and $\mathbf{b}$.
The calculated band structure of the basic structure
and interatomic distances in the modulated crystal
structure both indicate the aluminum atoms as
location of the CDW.
%, and they point towards
%Fermi surface nesting as prime mechanism.
%
The temperature dependence of the specific heat
reveals an anomaly at $T_{CDW} = 145$\ K
of a magnitude
similar to canonical CDW systems.
The present discovery of orthorhombic symmetry for
the CDW state of EuAl$_4$ leads to the suggestion
of monoclinic instead of orthorhombic symmetry
for the third AFM state.
\end{abstract}

     %-------------------------------------------------------------------------
     % The main body of the paper
     %-------------------------------------------------------------------------
     % Now enter the text of the document in multiple \section's, \subsection's
     % and \subsubsection's as required.

\section{\label{sec:eual4_introduction}%
Introduction}

\begin{figure}
\includegraphics[width=40mm]{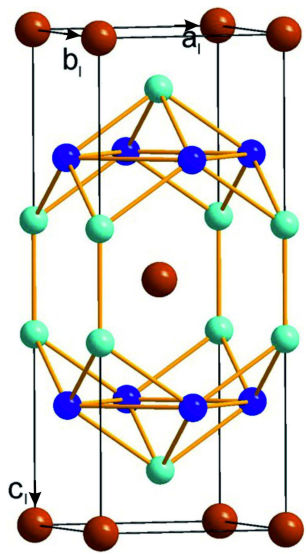}
\caption{Crystal structure of EuAl$_4$ with
space group $I4/mmm$ in the periodic phase at 250K.
Depicted is the $I$-centered unit cell with basis vectors
$\mathbf{a}_I$, $\mathbf{b}_I$ and $\mathbf{c}_I$.
Brown spheres correspond to the Eu atoms;
dark blue spheres represent Al1 atoms;
and green spheres stand for Al2 atoms.
Shortest interatomic distances are:
$d$[Eu--Eu] = 4.3949(2) \AA{},
$d$[Al1--Al1] = 3.1077(1) \AA{},
$d$[Al2--Al1] = 2.664(1) \AA{} and
$d$[Al2--Al2] = 2.568(4) \AA{}.}
\label{fig:eual4_basic_unit_cell}%
\end{figure}

EuAl$_4$ has attracted attention, because it develops
a charge-density wave (CDW) below $T_{CDW} = 145.1$\ K
and it exhibits four successive magnetic transitions
below 16\ K \cite{nakamura2015a,shimomura2019a}.
EuAl$_4$ adopts the BaAl$_4$ structure type that
has symmetry according to the tetragonal space
group $I4/mmm$ \cite{parth1983a, nakamura2015a},
as shown in Fig. \ref{fig:eual4_basic_unit_cell}.
It belongs to a large family of isostructural compounds,
including magnetic EuGa$_4$, fully ordered
EuAl$_2$Ga$_2$ and non-magnetic
SrAl$_4$ and BaAl$_4$
\cite{nakamura2016a,stavinoham2018a,onuki2020a}.
Recently, a symmetry-protected
non-trivial topology of the electronic
band structure was proposed for BaAl$_4$ \cite{wangk2021a}.
In case of magnetic EuAl$_4$, a chiral spin
structure, like skyrmions reported in
some divalent Eu compounds such as EuPtSi
\cite{Kaneko2019a,onuki2020a},
was proposed on the basis of the
observation of the topological
Hall resistivity and muon-spin rotation and
relaxation ($\mu SR$) studies \cite{shang2021a,zhuxy2022a}.
If such a nontrivial texture would be confirmed,
EuAl$_4$ would represent a rare case
of a compound where one could observe the
coexistence of exotic magnetic order and a CDW.
Since these exotic electronic and spin structures
depend on the symmetry,
knowledge of the true symmetry of the crystal
structure thus is of utmost importance for
understanding non-trivial magnetic properties.
Here we show that the CDW transition of EuAl$_4$
is accompanied
by a lowering of the symmetry towards orthorhombic,
and we present the incommensurately modulated CDW
crystal structure.

EuAl$_4$ is one of a few compounds \cite{schuttewj1993b},
where the lowering of the crystal symmetry
at a phase transition is
governed by the symmetry of the incommensurate
modulation wave describing the CDW, while any
lattice distortion could not be detected in the
present high-resolution diffraction experiment
with synchrotron radiation.
This feature might explain why the lowering
of symmetry has not been found in earlier studies
on EuAl$_4$.
% strong lattice distortions
This behavior is in contrast to other CDW
materials, like
Er$_2$Ir$_3$Si$_5$, Lu$_2$Ir$_3$Si$_5$ and
BaFe$_2$Al$_9$, for which the CDW transitions
are accompanied by large lattice distortions
\cite{ramakrishnan2020a,ramakrishnan2021a,meierwr2021a}
% CDW materials and mechanisms

% CDW in EuAl$_4$.
The phenomenon of CDW was originally identified
as a property of crystals with
quasi-one-dimensional (1D) electron bands,
such as NbSe$_3$ and K$_{0.3}$MoO$_3$
\cite{grunerg1994, monceaup2012a}.
A CDW is formed due to Fermi surface nesting (FSN),
where the nesting vector of the periodic structure
becomes the wave vector of the CDW of the
metallic bands as well as
of the accompanying modulation of the atomic
positions (periodic lattice distortion---PLD).
%$\mathbf{q}$ = $2\mathbf{k_{F}$
The modulation wave vector can be commensurate
or incommensurate with respect to the underlying
periodic basic structure.
More recent research has found that CDWs can
develop in crystalline materials that lack the
1D property of their crystal structures and
rather possess three-dimensionally (3D) structured
electron bands.
Alternate mechanisms have been proposed for
the formation of CDWs in 3D compounds, including
the mechanism of $q$-dependent electron-phonon
coupling (EPC)
\cite{zhux2015a,zhux2017a}.
Strongly correlated electron systems may also
support the formation of CDWs \cite{chencw2016a}.
The latter are often found for rare-earth
containing intermetallic compounds,
like the series of isostructural compounds
$R_5$Ir$_4$Si$_{10}$ ($R$ = rare earth)
\cite{ramakrishnan2017a}.

% Interplay magnetism and CDW
The interplay between CDWs and magnetism in
rare-earth compounds continues to attract
attention.
A competition between these two symmetry-breaking
phenomena can be expected, because both CDWs and
magnetic order depend on the Fermi surface through
FSN and the Ruderman-Kittel-Kasuya-Yosida (RKKY)
interaction between the localized magnetic moments
of the $4f$ electrons, as it is found for
EuAl$_4$ \cite{kobata2016a}.
In case of magnetoelastic coupling, the lattice
distortion may couple to the PLD or the lattice
distortion in the CDW state or to the EPC.
Experimentally, the coexistence of a CDW
and antiferromagnetic (AFM) order has been
established for
Er$_5$Ir$_4$Si$_{10}$ and
Sm$_2$Ru$_3$Ge$_{5}$
\cite{gallif2002a,kuo2020a}.
The series of compounds
$R$NiC$_2$ ($R$ = Pr, Nd, Gd, Tb, Dy, Ho, Er)
exhibit both CDW and AFM phase transitions
\cite{romanm2018a,shimomuras2016a, kolinciokk2017a,maeda2019a}.
SmNiC$_2$ is an exception in this series,
since it
develops ferromagnetic (FM) order below
$T_C = 17.7$\ K at which transition the CDW
is destroyed \cite{shimomura2009a,wolfel2010a}.
Recently, coexistence of CDW and FM orders
was found for the field-induced FM state of TmNiC$_2$ \cite{kolincio2020a}.

% CDW and Magnetism in EuAl$_4$
Magnetism of EuAl$_4$ is related to
localized magnetic moments
of the europium atoms in their divalent state:
the electronic configuration $4f^7$ implies
$J = S = 7/2$ and $L = 0$, where $J$ is the
total angular momentum, $S$ is the spin angular
momentum and $L$ is the orbital angular momentum
\cite{wernickjh1967a,nakamura2015a}.
This allows the study of the collective
magnetism without single-ion anisotropy,
as divalent Eu has zero orbital angular momentum.
Magnetic interactions are governed by
the RKKY
%Ruderman–Kittel–Kasuya–Yosida (RKKY)
interaction \cite{nakamura2015a}.
Neutron diffraction has established that
the AFM order involves an incommensurate
modulation wave \cite{kaneko2021a}.
Single-crystal x-ray diffraction (SXRD)
has shown the coexistence of the incommensurate
CDW modulation and AFM order \cite{shimomura2019a}.
Furthermore,
%Shimomura \textit{et al.}
\citeasnoun{shimomura2019a}
proposed a lowering of the lattice symmetry at
the third magnetic transition towards $Immm$
orthorhombic.
This is essentially different from the present
discovery of $Fmmm$ orthorhombic symmetry
below the CDW transition.

EuAl$_4$ possesses a 3D band structure with
localized $4f$ electrons of Eu well below the Fermi
surface \cite{kobata2016a}.
The CDW mainly involves orbitals of the Al atoms
\cite{kaneko2021a}.
This is in agreement with the observation of
a CDW with $T_{CDW}$ = 243 K in isostructural
SrAl$_4$, where non-magnetic Sr replaces Eu
\cite{nakamura2016a, haruo2020a}.
Here we present electronic band-structure calculations
for the tetragonal structure of EuAl$_4$.
They confirm that the location of the CDW is
on the Al atoms.
They reveal a 3D band structure with a
highly structured Fermi surface.
%featuring co-parallel planes on $k_z=0$ and $\pi$.
%This Fermi surface could be in agreement with
%FSN as mechanism for the formation of the CDW.

\section{\label{sec:eual4_experimental}%
Experimental and Computational details}

Single-crystals of EuAl$_4$ were synthesized by
the Al self-flux method.
The elements europium (Lieco, 99.9\% purity)
and aluminum (Alfa Aesar, 99.999\%) were filled
into an alumina crucible in the ratio 1:20.
The crucible was sealed in an evacuated quartz
glass ampoule.
It was heated to a temperature of 1323 K
and held at this temperature for 24 hours.
After which the crucible was cooled down to
1073 K in 6 hours and then slowly cooled at a
rate of 2 K/hr down to 923 K at which point the
crystals were separated from the molten metal
by centrifugation.
%\subsection{Chemical composition}
The 1:4 stoichiometry of the product was confirmed
by energy-dispersive x-ray spectroscopy (EDX) as
well as by the structure refinements against SXRD
data.

%\subsection{x-ray diffraction}
X-ray diffraction experiments were performed
at Beamline P24 of PETRA-III at DESY in Hamburg,
employing radiation of a wavelength of 0.5000 \AA{}.
The temperature of the specimen was controlled
by a CRYOCOOL open-flow helium gas cryostat.
Complete data sets of intensities of Bragg
reflections were measured at temperatures
of 250 K (tetragonal phase),
and of 70 K and 20 K (CDW phase).
Each run of data collection comprises $3640$ frames,
corresponding to a rotation of the crystal over
364 deg, which was repeated 10 times.
These data were binned to a data set of
364 frames of 1 deg of rotation and
10 seconds exposure time,
using the SNBL toolbox \cite{dyadkinv2016a}.
See Section S1 in the supporting information.

%\subsection{\label{sec:eual4_data_integration}%
%Data Integration}
%
The EVAL15 software suite \cite{schreursamm2010a}
has been used for processing the SXRD data.
At 250 K a single run was collected at a
crystal-to-detector distance of at 110 mm
and without a $2\theta$ offset of the detector.
At 70 K and at 20 K a crystal-to-detector distance
of 260 mm required two runs,
with and without $2\theta$ offset, respectively.
The two binned runs for 70 K and those for 20 K
were integrated separately, and subsequently
merged in the module ANY of EVAL15.
SADABS \cite{sheldrick2008}
has been used for scaling and absorption correction
with Laue symmetry $4/mmm$ for the 250 K data
and $mmm$ for 70 K and 20 K.
The reflection file produced was imported
to Jana2006 \cite{petricekv2014a, petricekv2016a}.
Table \ref{tab:eual4 crystalinfo}
shows the crystallographic information.

%\subsection{Magnetic susceptibility measurements}
The magnetic susceptibility $\chi(T)$ has been measured for temperatures 2.5--300\ K, using a commercial SQUID
magnetometer (MPMS5 by Quantum Design, USA). Measurements have been made in fields of 0.1 T and 0.5 T.

%\subsection{Specific heat measurements}
The specific heat, $C_{p}(T)$, has been measured
from 220 to 8 K by the thermal relaxation method,
using a
physical property measuring system (PPMS, Quantum Design, USA).

Density functional theory (DFT) based calculations
were performed within the generalized gradient
approximation (GGA) using the projector augmented
(PAW) wave method as implemented in the Vienna
{\it Ab-initio} Simulation Package (VASP)
\cite{Kresse1999a, Kresse1996a}.
The Perdew-Burke-Ernzerhof (PBE) functional
was used to consider the exchange-correlation
effects \cite{Perdew1996a}.
An energy cutoff of 380 eV was used for the
plane-wave basis set, and a $\Gamma$-centered,
$9\times 9 \times 9$ $k$ mesh was employed for
the bulk Brillouin zone sampling.
Spin-orbit coupling effects were considered
in all the calculations.
We employed Eu$^{2+}$ by considering the
remaining $4f$ electrons as core electrons.
A tight-binding
Hamiltonian was generated to compute the
Fermi surface on a finer $k$ grid \cite{Marzari1996a}.
The FermiSurfer software package was used to
visualize the Fermi surface \cite{Kawamura2019a}.

\section{\label{sec:eual4_discussion}%
Discussion}

\subsection{\label{sec:eual4_analysis_cdw}%
Analysis of the CDW structure}

\begin{figure}
\includegraphics[width=80mm]{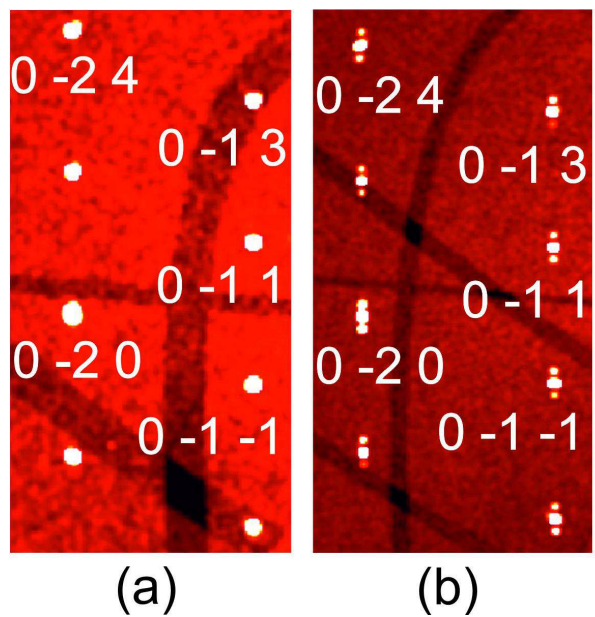}
\caption{Excerpt of the reconstructed reciprocal
layer $(0\,k\,l)$ for SXRD data measured at
(a) $T = 250$ K, and
(b) $70$ K.
Indices are given for several main reflections.
Panel (b) is better resolved than panel (a),
because of the longer crystal-to-detector distance
at 70 K.
Dark bands are due to insensitive pixels between
the active modules of the PILATUS3 X CdTe 1M detector.
}
\label{fig:eual4_unwarp}%
\end{figure}

SXRD at 250 K confirmed the $I4/mmm$
crystal structure of EuAl$_4$.
The SXRD data at 70 K revealed satellite
reflections at positions that can be
described by the modulation wave vector
$\mathbf{q} = (0, 0, 0.1781(3))$,
in agreement with the results by
\citeasnoun{nakamura2015a}
and \citeasnoun{shimomura2019a}.
Visualisation of the SXRD data was done with
aid of the software CrysAlisPRO \cite{crysalis}.
Figs. \ref{fig:eual4_unwarp}(a) and
\ref{fig:eual4_unwarp}(b)
show a small part of the $(0, k, l)$ plane of
the reconstructed reciprocal lattice.
Satellite reflections along $\mathbf{c}^{*}$
are clearly visible at 70 K
[Fig. \ref{fig:eual4_unwarp}(b)].
Upon further cooling to 20 K, there is
a reduction by 0.004 in the $\sigma_3$
component of
the modulation wave vector, in agreement
with \citeasnoun{shimomura2019a}.

We note that the lattice parameters in the
CDW phase do not give evidence for an
orthorhombic lattice distortion.
This could explain why earlier works
have not found this symmetry lowering.

In order to determine the crystal structure
of the CDW phase, we have tested different
superspace groups for its symmetry
(See Table S2 in the supporting information).
It is noticed that the tetragonal lattice
allows two fundamentally different orthorhombic
lattices as subgroups: $Immm$ preserves the
mirror planes perpendicular to the $\mathbf{a}$
and $\mathbf{b}$ axes of $I4/mmm$,
while $Fmmm$ preserves the diagonal mirror
planes.
Table \ref{tab:eual4_crystal_data}
provides the crystallographic data for three
of the refined crystal structures
(compare Table S2 in the supporting information).
The three models, A, B and C, are discussed below.

\subsubsection{Model A:}
In the diffraction pattern we did not
observe any splitting of the main or satellite reflections, where split reflections would
indicate a twinned crystal of lower symmetry.
Also, we did not find a distortion
in the lattice parameters, as it would occur
for a single-domain crystal of lower symmetry.
Furthermore, the preservation of tetragonal
symmetry within the CDW phase was reported in
the literature \cite{nakamura2015a, shimomura2019a}. Therefore, initial data processing was performed
under the assumption of tetragonal symmetry,
employing point group $4/mmm$ for scaling and
absorption correction in SADABS \cite{sheldrick2008}).
Structure refinements of the incommensurately
modulated structure were performed with a model
with superspace group $I4/mmm(00\sigma)0000$.
Table \ref{tab:eual4_crystal_data} shows that
$R_{int}$ as well as $R_F$ for the main reflections
are reasonably low, indicating that the average
structure of the CDW phase still is tetragonal
in good approximation.
This conclusion is reinforced by the fact that
refinement of the average structure against
main reflections leads to $R_F = 1.55\,\%$;
an excellent fit.
However, $R_F = 60.46\,\%$ for the satellite
reflections.
This high value indicates that the satellite
reflections are not well fitted and that the
CDW modulation does not have tetragonal symmetry.

This makes Model A an unsuitable candidate for the incommensurate CDW structure.
Other tetragonal superspace groups were also
tested, leading to similar failures in describing
the modulation wave or with reflection conditions
violated by the measured SXRD data
(see Table S2 in the supporting information).
As a result we can rule out tetragonal
symmetry for the modulated CDW crystal structure.

\subsubsection{Model B:}
As second model we have considered a lowering
of the symmetry from tetragonal $I4/mmm$ to its
orthorhombic subgroup $Immm$.
This orthorhombic point symmetry was used for
scaling and absorption correction of the SXRD
data in SADABS \cite{sheldrick2008}).
The CDW phase transition allows for
pseudomerohedral twinning of two,
differently oriented domains on the tetragonal
lattice, that are related by the missing fourfold
rotation \cite{parsons2003a}.
Since split reflections or a lattice distortion
could not be detected in the SXRD data,
all Bragg reflections have contributions from
both domains.
The structure refinement of a model in superspace
group $Immm(0 0 \sigma)s00$ has lead to a twin
volume ratio of $0.485 : 0.515$,
thus explaining the nearly tetragonal point symmetry
of the SXRD data.
$R$-values indicate a good fit to the SXRD data
for this model (Table \ref{tab:eual4_crystal_data}).
As a result, this model is a prime candidate for
describing the incommensurately modulated crystal
structure of the CDW phase.

\subsubsection{Model C:}
As last model we present model C with symmetry
according to $Fmmm$, the other orthorhombic subgroup,
which now preserves the diagonal mirror planes
of $I4/mmm$.
Scaling and absorption correction of the SXRD data
was performed with SADABS according to the
differently oriented point group $mmm$
\cite{sheldrick2008}.
Again, two domains are possible that are related
by the missing fourfold rotation.
The structure refinement of a model in superspace
group $Fmmm(0 0 \sigma)s00$ has lead to a twin
volume ratio of $0.454(4)\: :\: 0.546$,
thus explaining the nearly tetragonal point symmetry
of the SXRD data.
$R$-values indicate an excellent fit to the SXRD data
for this model (Table \ref{tab:eual4_crystal_data}),
which is significantly better than that of model B.
Furthermore, the refined parameters possess slightly
smaller s.u.'s in model C than in model B, while
the number of parameters is one smaller in model C
(Tables S3 and S4 in the supporting information).
Therefore, the best fit to the SXRD data
has been obtained for a modulated crystal structure
with symmetry according to the superspace group
$Fmmm(00\sigma)s00$.
$Fmmm(00\sigma)0s0$ is an alternate setting
of this superspace group, while all other
symmetries lead to a worse fit to the SXRD data
(Table S2 in the supporting information).

%Table \ref{tab:eual4 crystalinfo} shows the
%crystallographic data for the temperatures
%at 250 K, 70 K and 20 K.

\begin{figure}
\includegraphics[width=70mm]{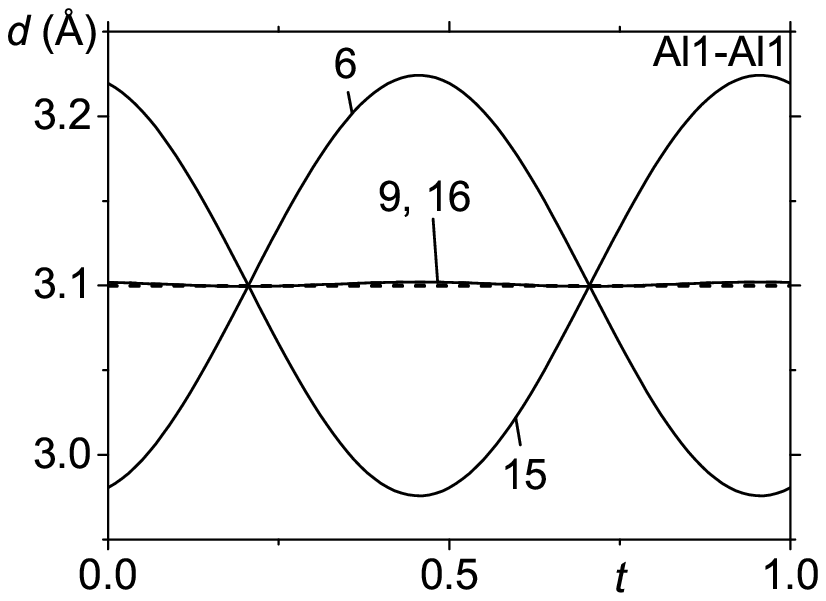}
\hfill
\includegraphics[width=70mm]{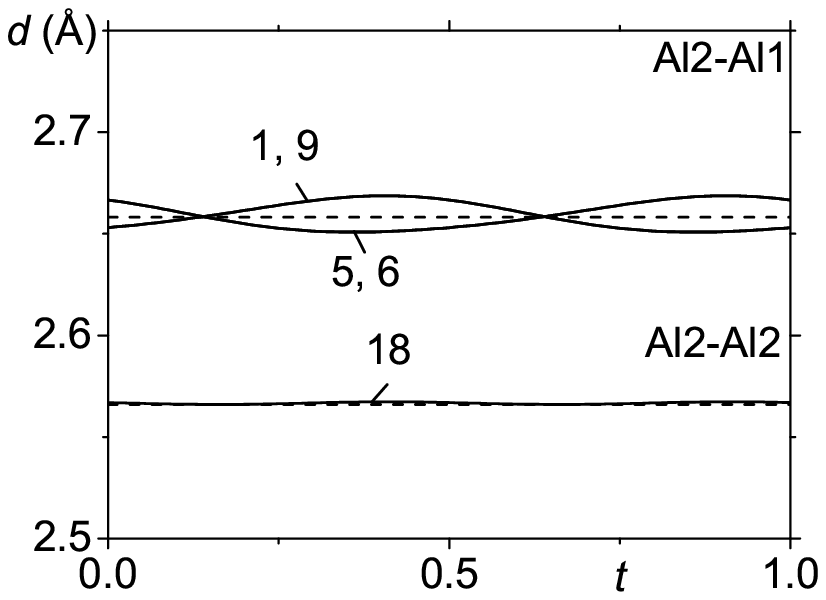}
\caption{$t$-Plot of interatomic distances (\AA{})
$d$[Al1--Al1], $d$[Al2--Al1] and $d$[Al2--Al2] at 70 K,
where the first atom is the central atom.
The number on each curve is the number of the symmetry
operator that is applied to the second atom of the
bond pair.
Symmetry operators are listed in Table S5
in the supporting information.}
\label{fig:eual4_t_plot_aluminium}%
\end{figure}

Recently, \citeasnoun{kaneko2021a} have proposed
that the CDW of EuAl$_4$ involves displacements
of the Al atoms perpendicular to $\mathbf{c}$,
while Eu would not be involved in the PLD.
The present crystal structure involves atomic
modulations exclusively perpendicular to $\mathbf{c}$,
as it is enforced by the superspace symmetry (Table S4).
However, the modulation amplitudes are of
comparable magnitude for all three atoms,
Eu, Al1 and Al2.
Nevertheless, a non-zero modulation amplitude is
not evidence by itself, that the involved atom
must contribute electronic states to the CDW.
The atomic modulation may also be caused by the
elastic coupling to other atoms that are carrying
the CDW.
In EuAl$_4$, the shortest interatomic distances are
between Al2 atoms and for Al1--Al2
[Fig. \ref{fig:eual4_t_plot_aluminium}(a)].
They are hardly modulated,
and forming a two-dimensional network of Al
perpendicular to $\mathbf{c}$
(Fig. \ref{fig:eual4_basic_unit_cell}).
The largest modulation is found for the next
shorter Al--Al distance between Al1 atoms
[Fig. \ref{fig:eual4_t_plot_aluminium}(b)].
This strong modulation suggests that the CDW
resides on the layers of Al atoms.
Eu is elastically coupled to Al1 and Al2
(Fig. S1) and is not part of the CDW.
The $t$ plots of interatomic distances cannot
elaborate on the precise location: the CDW
resides either on the Al1 atoms or on a
network of Al1 and Al2 atoms
(Fig. \ref{fig:eual4_t_plot_aluminium}).

\subsection{\label{sec:eual4_band_structure}%
Electronic structure and Fermi surface}

\begin{figure}
\includegraphics[width=120mm]{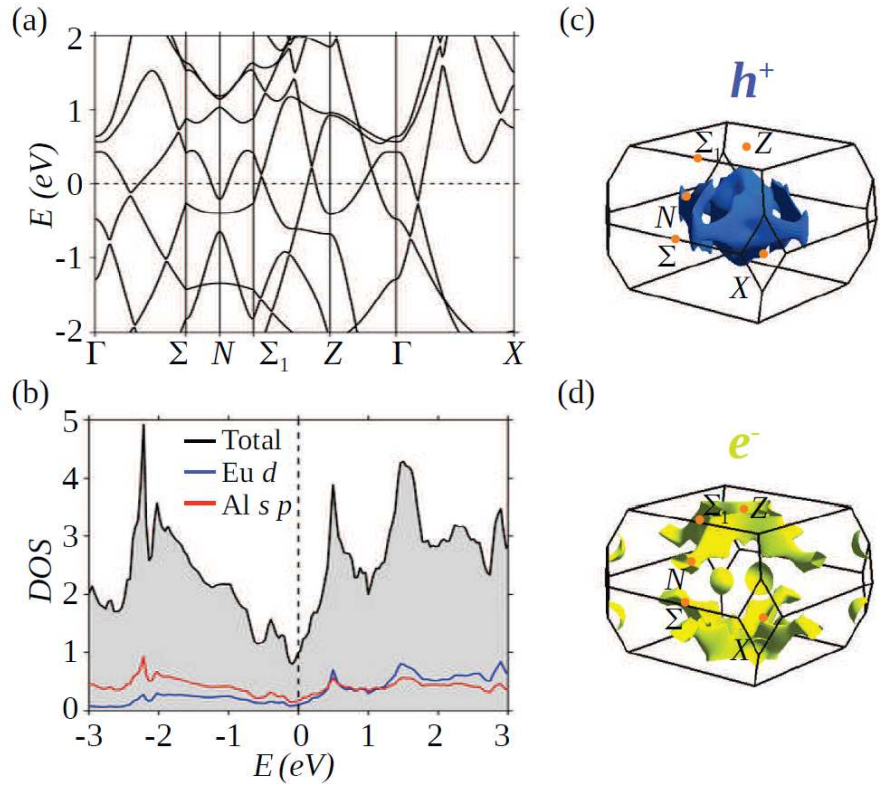}
\caption{(a) Bulk band structure and
(b) Density-of-states (DOS) of EuAl$_4$.
The dashed lines in (a) and (b) mark the Fermi level
at energy ($E$) zero.
The calculated
(c) hole (blue) and
(d) electron (yellow) Fermi pockets in
the primitive bulk Brillouin zone.
Three-dimensionally (3D) structured hole
and electron Fermi pockets are resolved.
%The corresponding cross-sectional
%Fermi contours at $k_z =0$ and $k_z=\pi $ planes are
%shown on the projected planes.
%The Fermi lines on cross-sectional planes resolve the %parallel sheets in the Fermi surface.
}
\label{fig:eual4_band_structure}%
\end{figure}

Figure \ref{fig:eual4_band_structure}(a) shows
the calculated band structure along the high-symmetry
directions in the primitive Brillouin zone
of the periodic crystal structure of EuAl$_4$
with $I4/mmm$ symmetry.
Both the valence and conduction bands cross
the Fermi level $E_F$, resolving its metallic
ground state.
Importantly, the bands along $\Gamma$--$Z$ have a
substantial energy dispersion.
This indicates the three-dimensional nature of
the Fermi surface as discussed below.
There is a Dirac nodal crossing above the Fermi
level along the $\Gamma$--$Z$ direction,
which is protected by $C_{4z}$ rotational symmetry.
EuAl$_4$ thus realizes a Dirac semimetallic state.
%Moreover, the states forming the Dirac nodes
%constitute a saddle-point singularity at the
%$\Gamma$-point, as illustrated with the dashed
%circle in Fig. \ref{fig:eual4_band_structure}(a).
To resolve the electronic states near $E_F$,
we present the atom-projected density of states (PDOS)
in Fig. \ref{fig:eual4_band_structure}(b);
Eu PDOS is in blue and Al PDOS is in red.
The Al states are dominant at $E_F$,
indicating that Al atoms are predominantly
metallic and more
likely undergo CDW modulations, in agreement
with the analysis of PLD
(Section \ref{sec:eual4_analysis_cdw})
and the literature.
PDOS of Eu comprises of $d$ states at the Fermi
level, while $4f$ states are well below $E_F$,
in agreement with the literature \cite{kobata2016a}.
%as found in our experiments similar
%to Ti atoms in TiSe$_2$ \cite{singh2017a}.

We present the calculated Fermi pockets
associated with the valence ($h^+$) and
conduction states ($e^-$) in Figs.
\ref{fig:eual4_band_structure}(c,d).
They reveal a hole pocket centered on $\Gamma$
and an electron Fermi pocket centered on $Z$.
%The associated cross-sectional Fermi contours
%are also shown at $k_z =0$ and $k_z=\pi$, respectively.
Both Fermi pockets are highly structured.
The $e^-$ Fermi pocket suggest the possibility
of nesting perpendicular to the $\mathbf{c}$ axes.
For the $h^+$ pocket a possible FSN is not clearly
resolved.
%which are clearly resolved in Fermi surface
%cuts on the $k_z =0$ and $k_z=\pi $ planes,
%respectively.
It should however be noted that because of the
3D nature of the Fermi surface, the nesting may
have a complicated structure.
These results do not clearly indicate FSN
as mechanism for the formation of
the CDW state in EuAl$_4$.

\subsection{\label{sec:eual4_magnetic_susceptibility}%
Magnetic susceptibility}

\begin{figure}
\includegraphics[width=80mm]{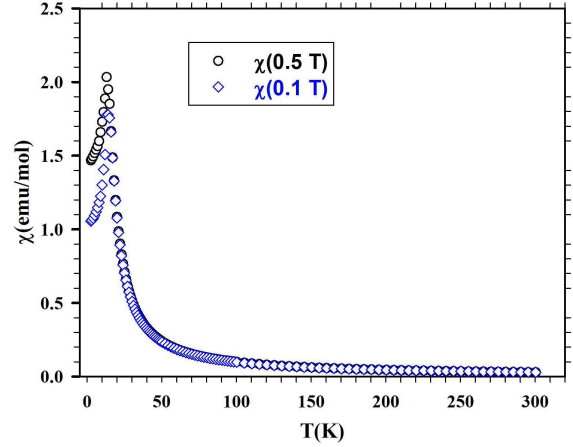}
\caption{Temperature dependent magnetic susceptibility
of EuAl$_4$ from 2.4 to 300 K.
Data measured in fields of 0.1 T and 0.5 T.}
\label{fig:eual4_magnetic_susceptibility}%
\end{figure}

The temperature dependence of the magnetic
susceptibility, measured with magnetic fields
of 0.1 T and 0.5 T, is shown for 2.4--300 K
in Fig. \ref{fig:eual4_magnetic_susceptibility}.
Any change of the susceptibility at the CDW transition
(\textit{e.g.} as a change of Pauli paramagnetism)
is masked by the large value of the paramagnetic
susceptibility.
However, the low temperature data reveal
an AFM transition below 16 K, which agrees with the previously published value \cite{nakamura2015a}.

\subsection{\label{sec:eual4_heat_capacity}%
Specific heat}

\begin{figure}
\includegraphics[width=80mm]{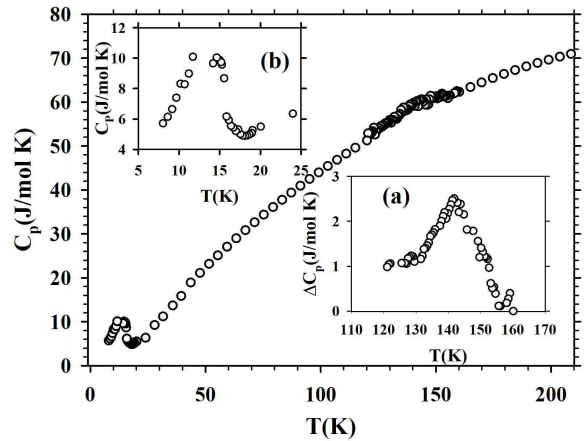}
\caption{Temperature dependence of the specific heat
$C_p$ from 8 to 210~K.
The lower inset (a) provides an enlarged view
around the anomaly at 145 K,
where $\Delta C$ = 2.5 J/(mol K).
The upper inset (b) displays $C_p$ vs $T$
at low temperatures 8--25 K.}
\label{fig:eual4_heat_capacity}%
\end{figure}

The temperature dependence of the specific heat
($C_p$) is shown for 8--210 K in
Fig. \ref{fig:eual4_heat_capacity}.
The high temperature data clearly
reveal a small, broad jump of
$\Delta C_p$ = 2.5 J/(mol K) at 145 K,
suggesting a thermodynamic
phase transition (CDW) at 145 K.
Such an anomaly in the temperature dependence
of the specific heat is consistent with that
observed in canonical CDW systems,
like NbSe$_3$ \cite{tomi1981a}.
The low-temperature data display multiple AFM
transitions, which have also been observed in
earlier studies \cite{nakamura2015a}.

\section{\label{sec:eual4_conclusions}%
Conclusions}

EuAl$_4$ possesses the BaAl$_4$ crystal structure type
with tetragonal symmetry $I4/mmm$.
It undergoes a CDW transition at $T_{CDW} = 145$\ K.
Here, we have presented the incommensurately
modulated crystal structure of EuAl$_4$ in its
CDW state.
Structure refinements according to the superspace
approach have shown that:
(i) the modulation is incommensurate with
modulation wave vector
$\mathbf{q} = (0, 0, 0.1781(3))$ at 70\ K,
in agreement with \citeasnoun{shimomura2019a};
and
(ii) the symmetry of the CDW crystal structure
is orthorhombic with superspace group
$Fmmm(00\sigma)s00$,
where $Fmmm$ is a subgroup of $I4/mmm$ of index 2.
Despite this group--subgroup relation, we did
not find any lattice distortion in the SXRD data.
Even more, atomic positions of the basic structure
of $Fmmm(00\sigma)s00$ still obey the $I4/mmm$
symmetry (Table S3).
Symmetry breaking is entirely in the modulation
wave (CDW and PLD), where atoms Eu and Al1
have displacements exclusively along $\mathbf{a}$ of the
$F$-centered unit cell (Table S4),
while the fourfold rotation
would require equal displacement amplitudes along
$\mathbf{a}$ and $\mathbf{b}$.

One interesting question is the location of the CDW.
Analysis of the modulation of interatomic distances
(Section \ref{sec:eual4_analysis_cdw})
as well as features of the electronic band structure
(Section \ref{sec:eual4_band_structure}) have
indicated the Al atoms as supporting the CDW,
in agreement with the literature
\cite{kobata2016a,kaneko2021a}.

Compounds of rare earth ($R$) and transition metals,
like $R_5$Ir$_4$Si$_{10}$ and $R_2$Ir$_3$Si$_5$,
contain highly correlated electron systems, with
accompanying influence on the mechanism of formation
of CDWs.
In EuAl$_4$, the majority element is the light $p$-block
metal aluminum.
Band-structure calculations have shown that FSN is
a possible mechanism of CDW formation.
Furthermore, the weak anomaly in $C_p(T)$ near
$T_{CDW}$ is similar to anomalies of canonical
CDW materials and it is much smaller than observed
for the rare-earth--transition-metal base compounds,
again this would support the FSN mechanism.
An alternative possibility is that the CDW is 
related to nesting of 
%Dirac points that are found on the $\Gamma$--$Z$ line  
nontrivial bands that are present 
in the band structure \cite{shiw2021a,chiuw2022a}.

The present discovery of orthorhombic symmetry for
the CDW state of EuAl$_4$ is important for modeling
of the electronic properties of the CDW state
as well as for identifying the correct magnetic
order and understanding the magnetic properties
of the four AFM states below 16 K.

Recent work either has used tetragonal symmetry
for analysing the AFM states
\cite{shang2021a, kaneko2021a}.
Alternatively, \citeasnoun{shimomura2019a} have
proposed orthorhombic $Immm$ symmetry for the
third AFM state.
This orthorhombic subgroup incorporates
the perpendicular mirror planes of $I4/mmm$,
while presently found $Fmmm$ is based on the
diagonal mirror planes of $I4/mmm$.
On the other hand, \citeasnoun{shimomura2019a}
report peak splitting in neutron diffraction
into ''three maxima.''
Together with the present observation of lowering
of symmetry at the CDW transition, this suggests
that the third AFM state could have monoclinic
symmetry ($\mathbf{c}$ unique)
instead of orthorhombic symmetry,
since, apparently, both the diagonal and
perpendicular mirror planes are lost.

     % Appendices appear after the main body of the text. They are prefixed by
     % a single \appendix declaration, and are then structured just like the
     % body text.

%\appendix
%\section{Appendix title}

     %-------------------------------------------------------------------------
     % The back matter of the paper - acknowledgements and references
     %-------------------------------------------------------------------------

     % Acknowledgements come after the appendices

\ack{Acknowledgements}
We acknowledge DESY (Hamburg, Germany),
a member of the Helmholtz Association HGF,
for the provision of experimental facilities.
Parts of this research were carried out at
PETRA III, using beamline P24.
Beamtime was allocated for proposal I-20190810.
J.‐K. Bao acknowledges the Alexander von Humboldt
Foundation for financial support.

     % References are at the end of the document, between \begin{references}
     % and \end{references} tags. Each reference is in a \reference entry.
\bibliographystyle{iucr}
\bibliography{eual4_cdw}%,sdw_cdw}

\clearpage

%---------------------------------------
% TABLES AND FIGURES SHOULD BE INSERTED
% AFTER THE MAIN BODY OF THE TEXT
%---------------------------------------

% Simple tables should use the tabular
% environment according to model

% Postscript figures can be included
% with multiple figure blocks

%
%
\begin{table}
\caption{Crystallographic data of crystal A of
EuAl$_4$ at 250 K, 70 K and 20 K.
Refinement method used: Least-squares on $F$.
The superspace group (SSG; No.) is given
according to \citeasnoun{stokesht2011a}.}
\label{tab:eual4 crystalinfo}%
\centering
\begin{tabular}{cccc}
Temperature (K) & 250 & 70 & 20 \\
\hline
Crystal system & Tetragonal & Orthorhombic & Orthorhombic \\
Space group; SSG & $I4/mmm$ & $Fmmm(00\sigma)s00$ &  $Fmmm(00\sigma)s00$\\
No. & 139 & {69.1.17.2} & {69.1.17.2}  \\
$a$ (\AA{}) &4.3949(1) &6.1992(4) & 6.1991(3) \\
$b$ (\AA{}) &4.3949    &6.2001(4)    & 6.1987(4) \\
$c$ (\AA{}) &11.1607(3)    &11.1477(3) & 11.1488(4) \\
Volume (\AA{}$^3$) & 215.57(1) & 428.47(4) & 428.41(4) \\
Wavevector, $q_z$ & - & 0.1781(3) & 0.1741(2) \\
$Z$ & 2 & 4 & 4 \\
Wavelength (\AA{}) & 0.50000 &0.50000 & 0.50000 \\
Detector distance (mm) &110 &260 & 260 \\
$2\theta$-offset (deg) &0 &0, 25 & 0, 25 \\
$\chi$-offset (deg) &-60 & -60 &-60 \\
Rotation per image (deg) & 1 & 1 & 1 \\
$(\sin(\theta)/\lambda)_{max}$ (\AA{}$^{-1}$) &0.682610& 0.748910 &0.749031 \\
Absorption, $\mu$ (mm$^{-1}$) & 5.8373 & 5.9090 & 5.9100 \\
T$_{min}$, T$_{max}$ & 0.3211, 0.3712 & 0.3209, 0.3732 &  0.3192, 0.3676 \\
Criterion of observability & $I>3\sigma(I)$ & $I>3\sigma(I)$ & $I>3\sigma(I)$ \\
Number of main reflections \\
measured &  1407 & 473 & 470 \\
unique (obs/all) & 109/109 &174/174 &176/176 \\
Number of satellites \\
measured & - & 929 &928 \\
unique (obs/all) & - & 279/316 & 263/322 \\
$R_{int}$ main (obs/all) &0.0374/0.0374 &0.0136/0.0136 &0.0188/0.0188 \\
$R_{int}$ sat (obs/all) &- &0.0581/0.0588 &0.0606/0.0616 \\
No. of parameters &9 &18 &18 \\
$R_{F }$ main (obs) &0.0147&0.0165 &0.0213 \\
$R_{F }$ sat (obs) &- &0.0369 &0.0311 \\
$wR_{F }$ main (all) &0.0214 &0.0203 &0.0230 \\
$wR_{F }$ sat (all) &- &0.0395 &0.0336 \\
$wR_{F }$ all (all) &0.0214 &0.0245 &0.0250 \\
GoF (obs/all) &1.53/1.53 &1.13/1.09 &0.93/0.88 \\
$\Delta\rho_{min}$, $\Delta\rho_{max}$(e \AA$^{-3}$) &
 -1.35, 1.15 &-2.40, 3.58 &-1.49, 1.58 \\
\end{tabular}
\end{table}
\begin{table}
\caption{Crystallographic data for three models
for the modulated crystal structure at 70 K,
based on different superspace groups.
Criterion of observability: $I>3\sigma(I)$}
\label{tab:eual4_crystal_data}%
\centering
\begin{tabular}{ccccc}
Model & A & B & C   \\
$a$ (\AA) &4.3834(3) &4.3835(3)  & 6.1992(4)    \\
$b$ (\AA) & 4.3834  &4.3841(3) & 6.2001(4)    \\
$c$ (\AA) &11.1488(4) & 11.1475(3) & 11.1477(3)  \\
$V$ (\AA$^3$) & 214.21(2) & 214.23(2)  & 428.47(4)  \\
\textbf{q}& 0.1782(3)\textbf{c}* & 0.1781(3)\textbf{c}*  & 0.1781(3)\textbf{c}* \\
SSG  &$I4/mmm(00\sigma)0000$ & $Immm(00\sigma)s00$ & $Fmmm(00\sigma)s00$ & \\
R$_{int}$ main (obs/all)$\%$  & 1.53/1.53 & 1.28/1.28  & 1.36/1.36  \\
R$_{int}$ sat (obs/all)$\%$  & 7.43/7.49 & 6.75/6.85  & 5.81/5.88  \\
R$_F$ main (obs/all)$\%$  & 4.37/4.37  & 1.96/1.96  & 1.65/1.65  \\
R$_F$ sat (obs/all)$\%$  & 60.46/70.53  & 5.25/5.83  & 3.69/4.05  \\
Unique main (obs/all)   & 130/130   & 225/225  &174/174  \\
Unique sat (obs/all)   & 215/254   & 365/425  &279/316  \\
No. of parameters  & 13  & 19  & 18   \\
\end{tabular}
\end{table}
%
%

%%%%%%%%%%%%%%%% F I G U R E S %%%%%%%%%
%
%

\end{document}